\documentclass[%
reprint,
 amsmath,amssymb,
 aps,
]{revtex4-2}
\usepackage{comment}
\usepackage{subcaption}
\usepackage{latexsym}
\usepackage{float}
\usepackage[usenames]{color}
\usepackage{colortbl}
\usepackage{graphicx}
\usepackage{dcolumn}
\usepackage{bm}
\usepackage{amsmath}
\begin{document}

\preprint{APS/123-QED}

\title{Collapses and revivals of polarization and radiation intensity induced by strong exciton-vibron coupling} 

\author{E. A. Tereshchenkov}
\affiliation{
 Dukhov Research Institute of Automatics (VNIIA), 22 Sushchevskaya, Moscow 127055, Russia;
}
\affiliation{
 Moscow Institute of Physics and Technology, 9 Institutskiy pereulok, Dolgoprudny 141700, Moscow region, Russia;
}
\affiliation{
 Institute for Theoretical and Applied Electromagnetics, 13 Izhorskaya, Moscow 125412, Russia;
}

\author{V. Yu. Shishkov}
\affiliation{
 Dukhov Research Institute of Automatics (VNIIA), 22 Sushchevskaya, Moscow 127055, Russia;
}
\affiliation{
 Moscow Institute of Physics and Technology, 9 Institutskiy pereulok, Dolgoprudny 141700, Moscow region, Russia;
}

\author{E. S. Andrianov}
\affiliation{
 Dukhov Research Institute of Automatics (VNIIA), 22 Sushchevskaya, Moscow 127055, Russia;
}
\affiliation{
 Moscow Institute of Physics and Technology, 9 Institutskiy pereulok, Dolgoprudny 141700, Moscow region, Russia;
}%

\date{\today}

\begin{abstract}
Recently, systems with strong coupling between electronic and vibrational degrees of freedom attract a great attention.
In this work, we consider the transient dynamics of the system consisting of strongly coupled vibron and exciton driven by external monochromatic field.
We show that under coherent pumping, polarization of exciton exhibits complex quantum dynamics which can be divided into three stages.
At the first stage, exciton oscillations at its eigenfrequency relax due to the transition to set of shifted Fock states of vibrons.
We demonstrate that these shifted Fock states play the role of an effective reservoir for the excited exciton state.
The time of relaxation to this reservoir depends on exciton-vibron coupling.
At the second stage, excitation, transferred to the reservoir of the vibronic shifted states at the first stage, returns into electronic degrees of freedom and revival of oscillations at exciton eigenfrequency appears.
Thus, the dynamics of molecular polarization exhibit collapses and revivals.
At the final stage, these collapses and revivals dissipate and polarization exhibits Rayleigh response at the frequency of the external field.
Discovered collapses and revivals manifest in radiation spectrum as multiple splitting of the spectral line near the exciton transition frequency.

\end{abstract}

\maketitle

\section{INTRODUCTION}\label{sec:introduction}
Interaction between electronic and nuclei vibration degrees of freedom was investigated for almost a hundred years since discovery of inelastic Raman scattering in molecules~\cite{raman1928new} and Brillouin scattering in solids~\cite{landsberg1928new}.
This interaction lies at the ground of many phenomena discovered from that time including coherent Raman scattering~\cite{long2002raman,sathyanarayana2015vibrational,schrader2008infrared,schrader2008infrared,weber2012raman}, Raman and Brillouin lasers~\cite{feng2017raman,islam2004overview,white1987stimulated}, optomechanics~\cite{schmidt2017linking,roelli2016molecular,lombardi2018pulsed}, polariton chemistry~\cite{herrera2017absorption,herrera2018theory}, room-temperature vibron-mediated Bose-Einstein condensate~\cite{zasedatelev2019room,zasedatelev2021single}.

First description of phenomena associated with this interaction was based on formalism of absorption and emission of quanta of nuclei vibrations, e.g., acoustical or optical phonons.
These absorption and emission processes take place during interaction of incident field with the electronic degrees of freedom~\cite{placzek1934rayleigh,long2002raman}, which parametrically interact with nuclei vibrations degrees of freedom.
Developments of theory of open quantum system~\cite{nakajima1958quantum,zwanzig1960ensemble,breuer2002theory,grabert2006projection} and solid-state theory~\cite{frohlich1954electrons,koppel1984multimode,born2000quantum}, enabled to give more deep understanding of these phenomena.
Namely, in the case when the external field interacts with electronic degrees of freedom only (which takes place for Raman-active molecules), the energy of the dipole moment excited by the coherent field results in shift of nuclei from equilibrium position and their subsequent oscillations~\cite{shishkov2019enhancement}.
Such interaction between electronic and nuclei degrees of freedom can be described by Frohlich~\cite{frohlich1954electrons,koppel1984multimode,born2000quantum}, or Holstein-Tavis~\cite{herrera2018theory, herrera2017absorption, herrera2016cavity, herrera2017dark} Hamiltonian.
The last describes effective shift of transition frequency of electronic degree of freedom proportional to the amplitude (generally, depending on time) of nuclei vibrations~\cite{shishkov2019enhancement}.
It has been demonstrated that this interaction is responsible for spontaneous Raman scattering~\cite{shishkov2019enhancement}, Raman lasers~\cite{shishkov2019connection}, surface-enhanced Raman scattering (SERS)~\cite{neuman2020quantum} and formation of third-order nonlinear susceptibility produced by nuclei vibrations~\cite{shishkov2019connection}.

Recently, the systems with strong coupling between electronic degrees of freedom, e.g., exciton, and nuclei vibration degrees of freedom, e.g., optical phonon or vibron, have been actively investigated~\cite{holmes2004strong,holmes2007strong,george2015ultra}.
The interaction between exciton and vibron can be described by the Frohlich constant of interaction $g$~\cite{kabuss2013threshold,shishkov2019enhancement,shishkov2019connection} or, alternatively, Huang-Rhys factor, $\lambda = g/\omega_{\rm{v}}$~\cite{herrera2018theory, herrera2017absorption, herrera2016cavity, herrera2017dark}.
The large value of Huang-Rhys factor, $\lambda \simeq 10$, is achieved in organic materials for high-frequency vibrational modes~\cite{holmes2004strong,holmes2007strong,george2015ultra}.

To take into account dissipation processes which inevitably present in any real molecules, modern theories describe the dynamics of coupled electronic and nuclei degrees of freedom excited by external field in terms of master equation for the density matrix~\cite{may2008charge}.
When reservoir degrees of freedom are eliminated in the Born-Markov approximation, such master equation has Lindblad form~\cite{lindblad1976generators,gorini1976completely}.
Besides Hermitian interaction, it takes into account dissipation processes.
Generally, to correctly describes the dynamics of open quantum system with the strong coupling between subsystems, it is necessary to find eigenstates of the interacting subsystems~\cite{vovchenko2021model,kosloff2013quantum}, and then consider relaxation-induced transition between these eigenstates.
This approach is called global approach to the master equation for the density matrix~\cite{kosloff2013quantum,vovchenko2021model}.
It is applied when the distances between eigenstates are larger than the dissipation rates~\cite{DaviesMME} which is fulfilled at least in the case of strong interaction between subsystems of an open quantum system~\cite{vovchenko2021model}.
In the opposite case one can use local approach which treats relaxation of interacting subsystems independently~\cite{shishkov2019relaxation}.

Recently, the dynamics of excitons strongly interacting with vibrons has been investigated in the context of phonon lasers~\cite{kabuss2012optically, kabuss2013threshold}, though in the local approach for the Lindblad master equation for the density matrix, or, with applying mean-field theory~\cite{shishkov2019connection}.
In such a problems, the attention has been paid to stationary state of the system.
However, the interest is also of transient dynamics of strongly coupled exciton-vibron system, in particular, for the problem with ultra-short pumping.

Here we consider the transient dynamics of strongly coupled exciton and vibron in the global approach.
We solve the Lindblad master equations using eigenstates of the Hamiltonian of strongly coupled exciton and vibron.
We show that the transient dynamics of the system can be divided into three stages. 
At the first stage, oscillations of the polarization at the exciton eigenfrequency collapse due to transition to the set of shifted Fock states of vibrons.
At the second stage, revivals of these oscillations appear.
We give theoretical estimates for the collapse and revival times and find the dependence of this time on the interaction constant between the vibron and exciton and the frequency of the vibron.
We demonstrate that these collapses and revivals lead to the radiation spectrum exhibits a series of peaks near the exciton transition frequency with a distance equal to vibron frequency.

\section{THE MODEL}\label{sec:model_description}
For simplicity, we assume that the exciton corresponding to the excitation of an electron-hole pair interacts with one eigenmode of oscillations of the nuclei, hereinafter, vibron.
Such a situation is realized, e.g., in semiconductor quantum dots~\cite{guzelturk2014amplified, hoogland2013optical, anantathanasarn2006lasing, heinrichsdorff1997room, ishikawa1998self}, and organic molecule~\cite{altman1991solid, peterson1970cw, duarte1990dye}.
Among the electronic states, we consider one state in the valence band $\left| g \right\rangle $, which we will call the ground state, and one state in the conduction band $\left| e \right\rangle $, which we will call the excited state.
We denote the exciton eigenfrequency as $\omega_{\sigma}$.
If we suppose that the energy of the ground state is zero, then the exciton Hamiltonian takes the form $\hbar {\omega _\sigma }\left| e \right\rangle \left\langle e\right|$.
Further, we introduce operators of transitions from the excited state $\left| e \right\rangle $ to the ground state $\left| g \right\rangle $, $\hat \sigma = \left| g \right\rangle \left\langle e \right|$ and operator of inverse transition, ${\hat \sigma ^\dag } = \left| e \right\rangle \left\langle g \right|$.
In such approximation, the exciton Hamiltonian can be written as $\hbar {\omega _\sigma }{\hat \sigma ^\dag }\hat \sigma $.
The operators $\hat \sigma $ and ${\hat \sigma ^\dag }$ satisfy the commutation relation $\left[ \hat\sigma^\dag, \hat\sigma \right] = \hat D$, where the operator $\hat D$ is the operator of the difference between the populations of the excited and ground states.
Note that the mean value of the operator $\hat \sigma$ multiplied by matrix element of dipole moment ${\bf{d}}_{eg}$ and molecular concentration gives molecular polarization.
Thus, the expected value of the operator $\hat \sigma$ can be called dimensionless polarization.

We also consider one vibrational mode of QD nuclei in the harmonic approximation.
In this case, the Hamiltonian can be represented as $\hbar {\omega_{\rm{v}}}{\hat b^\dag }\hat b$, where ${\omega _{\rm{v}}} $ is the eigenfrequency of the vibron, ${\hat b^\dag}$ and $\hat b$ are the creation and annihilation operators of vibrons satisfying the commutation relation $\left[ {\hat b,{{\hat b}^ \dag }} \right] = \hat 1$.
The interaction between the electronic and vibrational subsystems of molecules, i.e., the exciton and the vibron, can be represented in the form of Frohlich Hamiltonian $\sigma^{\dag}\sigma\left( \hat b^\dag + \hat b \right)$~\cite{may2008charge,herrera2016cavity,herrera2018theory}, where $g$ is the interaction constant.
The operator ${\hat b^\dag } + \hat b$ is the nuclear displacement amplitude operator, and the operator ${\hat \sigma ^\dag }\hat \sigma $ is the population operator of the excited state of the exciton.

The Hamiltonian of the system has the form
\begin{gather}
{\hat H_{\rm{mol}}} = \hbar {\omega _\sigma }{\hat \sigma}^\dag {\hat \sigma} + \hbar {\omega _{\rm{v}}}{\hat b}^\dag {\hat b} + \hbar g{\hat \sigma}^\dag {\hat \sigma}\left( {{{\hat b}}^\dag  + {{\hat b}}} \right)
\label{eq:Hamiltonian_mol}
\end{gather}

We will consider coherent pumping of exciton by monochromatic electromagnetic (EM) wave.
The EM wave is described in the classical approximation.
In the dipole approximation, the interaction Hamiltonian has the form
\begin{equation}
{\hat H_{\rm{mol - field}}} = \frac{{\hbar {\Omega}}}{2}\left( {{{\hat \sigma }}{e^{i\omega t}} + {{\hat \sigma }}^\dag {e^{ - i\omega t}}} \right)
\label{eq:Hamiltonian_mol-field}
\end{equation}
where $\omega$ is the EM field frequency, ${\Omega}$ is the Rabi interaction constant with the exciton dipole moment:
\begin{equation}
{\Omega} = \frac{{ - {{\bf{d}}_{eg}}{\bf{E}}({r})}}{\hbar}
\label{eq:Omega_Raby}
\end{equation}
Here ${{\bf{d}}_{eg}}$ is the transition matrix element between states $\left| e \right\rangle $ and $\left| g \right\rangle $, ${\bf{E}}({{\bf{r}}})$ is the electric field amplitude at the exciton location (we suppose that the length of exciton localization is much smaller than the EM wavelength).
The total Hamiltonian of the system reads
\begin{equation}
\begin{gathered}
\hat H = \hbar {\omega _\sigma }{{\hat \sigma}}^\dag {{\hat \sigma}} + \hbar {\omega _{\rm{v}}}{{\hat b}}^\dag {{\hat b}} + \hbar g{{\hat \sigma}}^\dag {{\hat \sigma}}\left( {{{\hat b}}^\dag + {{\hat b}}} \right)\\
+ \frac{{\hbar {\Omega}}}{2}\left( {{{\hat \sigma}}{e^{i\omega t}} + {{\hat \sigma}}^\dag {e^{ - i\omega t}}} \right)
\label{eq:Hamiltonian}
\end{gathered}
\end{equation}

To find an eigenstates, we note that the molecule Hamiltonian can be rewritten in the form
\begin{gather}\label{HamSys2}
	\hat H_{{\rm{mol}}} = \hbar \omega_{\sigma} \left( 1 - \alpha^2 \right) \hat \sigma^{\dag} \hat \sigma + \hbar \omega_{{\rm{v}}} \hat{\tilde b}^{\dag} \hat{\tilde b}
\end{gather}
where
\begin{gather}\label{SfiftOp}
	\hat{\tilde b} = \hat b +  \alpha \hat \sigma^{\dag} \hat \sigma, \quad \alpha = \frac{g}{\omega_{{\rm{v}}}}
\end{gather}
is shifted annihilation operator of vibrons.
The eigenstates of the Hamiltonian~(\ref{HamSys2}) have the form
\begin{gather}\label{ShiftStatesGround}
	\left| g, n \right\rangle, \omega_{gn} = \omega_{{\rm{v}}}\left(n+1/2\right), n=0,1,...
\end{gather}
\begin{gather}\label{ShiftStatesEx}
	\left|e, \tilde n_{\alpha} \right\rangle, \omega_{e\tilde n} = \omega_{\sigma}\left(1-\alpha^2\right) + \omega_{{\rm{v}}}\left(\tilde n+1/2\right), \tilde n = 0,1,..
\end{gather}
where we introduce shifted Fock state~\cite{may2008charge,al1969scattering} via displacement operator $\hat D\left(\alpha\right) = \exp\left(\alpha \hat b^{\dag} - \alpha^* \hat b\right)$:
\begin{equation}
\left|\tilde n_{\alpha}\right\rangle=\hat{D}(\alpha)\left|n\right\rangle
\label{eq:displaced_operator}
\end{equation}
The matrix elements of expansion of the shifted Fock states over the Fock state are well-known~\cite{al1969scattering,barnett2002methods}
\begin{gather}\label{MatrEl}
	\left\langle n_{0} | \tilde n_{\alpha} \right\rangle  = (-\alpha^{*})^{\tilde n_{\alpha}-n_{0}} \times
	\\ \nonumber
	\times \sqrt{\frac{n_{0}!}{\tilde n_{\alpha}!}}L_{n_{0}}^{(\tilde n_{\alpha}-n_{0})}\left(\left|\alpha\right|^{2}\right)e^{-\left|\alpha\right|^{2}/2},\quad {\rm{when}} \quad \tilde n_{\alpha} \geq n_{0}
	\\ \nonumber
\left\langle n_{0} | \tilde n_{\alpha} \right\rangle  =(\alpha^{*})^{n_{0}-\tilde n_{\alpha}}\times
	\\ \nonumber
	\sqrt{\frac{\tilde n_{\alpha}!}{n_{0}!}}L_{\tilde n_{\alpha}}^{(n_{0}-\tilde n_{\alpha})}\left(\left|-\alpha\right|^{2}\right)e^{-\left|-\alpha\right|^{2}/2},\quad {\rm{when}} \quad  \tilde n_{\alpha}<n_{0}
\end{gather}
and are related to Frank-Condon factors~\cite{may2008charge}.
Here $L_{m}^{n}$ are associated Laguerre polynomials.

To describe relaxation processes, we use the Lindblad equation for the density matrix ~\cite{scully1999quantum,carmichael1999statistical,breuer2002theory} in the global approach~\cite{kosloff2013quantum,shishkov2019relaxation}.
It can be derived from the von Neumann equation for the density matrix of the system and environment by eliminating environmental variables in the Born-Markov approximation~\cite{scully1999quantum, carmichael1999statistical, breuer2002theory}.
The Lindblad equation has the form
\begin{gather}
\dot{\hat{\rho}}  = -\frac{i}{\hbar }[\hat H,\hat \rho ] + {L_{\rm{diss}}}[\hat \rho ] + {L_{\rm{pump}}}[\hat \rho ] + {L_{\rm{deph}}}[\hat \rho ] + {L_{\rm{v}}}[\hat \rho ]
\label{eq:Lindblad_equation_first}
\end{gather}
The first term on the right side of the equation (\ref{eq:Lindblad_equation_first}) describes the Hermitian dynamics of the system.
The rest terms describe the relaxation processes that occur when the system interacts with the environment (external reservoirs).

For further consideration, it is convenient to expand the density matrix over the eigenstates~(\ref{ShiftStatesGround}),(\ref{ShiftStatesEx}):
\begin{equation}
\hat \rho = \hat \rho_{gg} + \hat \rho_{eg} + \hat \rho_{ge} + \hat \rho_{ee}
\label{eq:rho}
\end{equation}

where
\begin{equation}
\begin{gathered}
\hat \rho_{gg} = \sum\limits_{n_{1}, n_{2}} \rho_{g, n_{1}, g, n_{2}}\left| g, n_{1} \right\rangle \left\langle g, n_{2}\right|\\
\hat \rho_{eg} = \sum\limits_{\tilde{n}_{1\alpha}, n_{2}} \rho_{e, \tilde{n}_{1\alpha}, g, n_{2}}\left| e, \tilde{n}_{1\alpha} \right\rangle \left\langle g, n_{2}\right|\\
\hat \rho_{ge} = \sum\limits_{n_{1}, \tilde{n}_{2 \alpha}} \rho_{g, n_{1}, e, \tilde{n}_{2 \alpha}}\left| g, n_{1} \right\rangle \left\langle e, \tilde{n}_{2\alpha}\right|\\
\hat \rho_{ee} = \sum\limits_{\tilde{n}_{1\alpha}, \tilde{n}_{2\alpha}} \rho_{e, \tilde{n}_{1\alpha}, e, \tilde{n}_{2\alpha}}\left| e, \tilde{n}_{1\alpha} \right\rangle \left\langle e, \tilde{n}_{2\alpha}\right|
\end{gathered}
\label{eq:rho_detail}
\end{equation}
where $n$ is the number of vibrons in the ground state, $\tilde{n}$ is the number of "shifted vibrons" in the excited state.

The Lindblad superoperators describing the relaxation of vibrons have the form
\begin{equation}
\begin{gathered} \label{eq:Lindblad_v}
{L_{\rm{v}}}[\hat \rho ] = {L_{\rm{v}}}[\hat \rho_{gg} ] + {L_{\rm{v}}}[\hat \rho_{ee} ]\\
	{L_{\rm{v}}}[\hat \rho_{gg} ] = \frac{{{\gamma _{\rm{v}}(1+n_{{\rm{v}}})}}}{2}\left( {2{{\hat b}}\hat \rho_{gg} \hat b^\dag  - \hat b^\dag {{\hat b}}\hat \rho_{gg}  - \hat \rho_{gg} \hat b^\dag {{\hat b}}} \right)\\
	+ \frac{{{\gamma _{\rm{v}}n_{{\rm{v}}}}}}{2}\left( {2{{\hat b^\dag}}\hat \rho_{gg} \hat b  - \hat b {{\hat b^\dag}}\hat \rho_{gg}  - \hat \rho_{gg} \hat b {{\hat b^\dag}}} \right)\\
{L_{\rm{v}}}[\hat \rho_{ee} ] = \frac{{{\gamma _{\rm{v}}(1+n_{{\rm{v}}})}}}{2}\left( {2{{\hat{\tilde{b}}}}\hat \rho_{ee} \hat{\tilde{b}}^\dag  - \hat{\tilde{b}}^\dag {{\hat{\tilde{b}}}}\hat \rho_{ee}  - \hat \rho_{ee} \hat{\tilde{b}}^\dag {{\hat{\tilde{b}}}}} \right)\\
+ \frac{{{\gamma _{\rm{v}}n_{{\rm{v}}}}}}{2}\left( {2{{\hat{\tilde{b}}^\dag}}\hat \rho_{ee} \hat{\tilde{b}}  - \hat{\tilde{b}} {{\hat{\tilde{b}}^\dag}}\hat \rho_{ee}  - \hat \rho_{ee} \hat{\tilde{b}} {{\hat{\tilde{b}}^\dag}}} \right)
\end{gathered}
\end{equation}
The term $L_{\rm{v}}\left[\rho_{gg}\right]$ describes the transitions between the levels $\left|e,\tilde n\right>$ with different $\tilde n$.
The term $L_{\rm{v}}\left[\rho_{ee}\right]$ describes the transitions between the levels $\left|g,n\right>$.
Here $n_{\rm{v}} = \left(\exp\left(\hbar\omega_{\rm{v}}/kT\right) - 1\right)^{-1}$ is the mean number of quanta in the reservoir taken at the phonon frequency.
The ratio of transition rates with energy increasing to energy lowering is $\gamma_{\rm{v}} n_{\rm{v}} / \gamma_{\rm{v}}\left(1+n_{\rm{v}}\right)= \exp\left(-\hbar\omega_{\rm{v}}/kT\right)$, i.e., satisfies Kubo-Martin-Schwinger condition~\cite{breuer2002theory}.

The Lindblad superoperator describing phase destruction (dephasing) of exciton is
\begin{equation}
{L_{\rm{deph}}}[\hat \rho ] = \frac{{{\gamma _{\rm{deph}}}}}{4}\left( {{{\hat D}}\hat \rho {{\hat D}} - \hat \rho } \right)
\label{eq:Lindblad_deph}
\end{equation}
The Lindblad superoperator, which describes the relaxation of the exciton energy, takes the form
\begin{equation}
	{L_{\rm{diss}}}[\hat \rho ] = \frac{{{\gamma _{D}}}}{2}\sum_{\Delta \omega}\left( 2\hat {\tilde \sigma}_{\Delta \omega} \hat \rho \hat {\tilde \sigma}^\dag_{\Delta \omega}  - \hat {\tilde \sigma}^\dag_{\Delta \omega} {{\hat {\tilde \sigma}_{\Delta \omega} }}\hat \rho  - \hat \rho \hat {\tilde \sigma}^\dag_{\Delta \omega} {{\hat {\tilde \sigma} }} \right)
\label{eq:Lindblad_diss}
\end{equation}
where $\Delta \omega = \omega_{e \tilde n} - \omega_{gn}$ and $\hat {\tilde \sigma}_{\Delta \omega} = \left< g, n \right| \hat \sigma \left| e, \tilde n\right> \left|g,n\right> \left<e,\tilde n\right|$.
The Lindblad superoperator describing incoherent exciton pumping is
\begin{equation}
{L_{\rm{pump}}}[\hat \rho ] = \frac{{{\gamma _{p}}}}{2}\sum_{\Delta \omega}\left( 2\hat {\tilde \sigma}_{\Delta \omega} \hat \rho \hat {\tilde \sigma}^\dag_{\Delta \omega}  - \hat {\tilde \sigma}^\dag_{\Delta \omega} {{\hat {\tilde \sigma}_{\Delta \omega} }}\hat \rho  - \hat \rho \hat {\tilde \sigma}^\dag_{\Delta \omega} {{\hat {\tilde \sigma} }} \right)
\label{eq:Lindblad_pump}
\end{equation}

Using (\ref{eq:Lindblad_equation_first}), we can write an equation for each matrix element of (\ref{eq:rho}) 
\begin{equation}
\begin{gathered}
	\dot\rho_{e, \tilde{n}_{1\alpha}, e, \tilde{n}_{2\alpha}}=i\omega_{{\rm{v}}}(\tilde{n}_{1\alpha}-\tilde{n}_{2\alpha})\rho_{e, \tilde{n}_{1\alpha}, e, \tilde{n}_{2\alpha}}\\
+i\Omega e^{-i\omega t}\sum\limits_{n_{1}} \left\langle\tilde{n}_{1\alpha}|n_{1}\right\rangle\rho_{g, n_{1}, e, \tilde{n}_{2\alpha}}\\
-i\Omega e^{i\omega t}\sum\limits_{n_{2}} \left\langle n_{2}|\tilde{n}_{2\alpha}\right\rangle\rho_{e, \tilde{n}_{1\alpha}, g, n_{2}}\\
-\gamma_{D}\rho_{e, \tilde{n}_{1\alpha}, e, \tilde{n}_{2\alpha}} 
+ \gamma_{p}\sum\limits_{n_{1}, n_{2}} \left\langle n_{1}|\tilde{n}_{1\alpha}\right\rangle \left\langle n_{2}|\tilde{n}_{2\alpha}\right\rangle \rho_{g, n_{1}, g, n_{2}}\\
	+\frac{\gamma_{{\rm{v}}}(\tilde{n}_{0}+1)}{2}\left(2\sqrt{(\tilde{n}_{1\alpha}+1)(\tilde{n}_{2\alpha}+1)}\rho_{e, \tilde{n}_{1\alpha}+1, e, \tilde{n}_{2\alpha}+1}\right.\\
-(\tilde{n}_{1\alpha}+\tilde{n}_{2\alpha})\rho_{e, \tilde{n}_{1\alpha}, e, \tilde{n}_{2\alpha}}\Bigr)\\
	+\frac{\gamma_{{\rm{v}}}\tilde{n}_{0}}{2}\left(2\sqrt{\tilde{n}_{1\alpha}\tilde{n}_{\alpha2}}\rho_{e, \tilde{n}_{1\alpha}-1, e, \tilde{n}_{2\alpha}-1}\right.\\
-(\tilde{n}_{1\alpha}+\tilde{n}_{2\alpha}+2)\rho_{e, \tilde{n}_{1\alpha}, e, \tilde{n}_{2\alpha}}\Bigr)
\end{gathered}
\label{eq:dot_rho_ee}
\end{equation}
\begin{equation}
\begin{gathered}
	\dot\rho_{g, n, e, \tilde{n}_{\alpha}}=i\omega_{\sigma}\rho_{g, n, e, \tilde{n}_{\alpha}}+i\omega_{{\rm{v}}}(\tilde{n}_{\alpha}-n)\rho_{g, n, e, \tilde{n}_{\alpha}}\\
+i\Omega e^{i\omega t}\sum\limits_{\tilde{n}_{1\alpha}} \left\langle n|\tilde{n}_{1\alpha}\right\rangle\rho_{e, \tilde{n}_{1\alpha}, e, \tilde{n}_{\alpha}}\\
-i\Omega e^{i\omega t}\sum\limits_{n_{2\alpha}} \left\langle n_{2}|\tilde{n}_{\alpha}\right\rangle\rho_{g, n, g, n_{2}}\\
-\frac{\gamma_{p}+\gamma_{D}+\gamma_{deph}}{2}\rho_{g, n, e, \tilde{n}_{\alpha}}
\end{gathered}
\label{eq:dot_rho_ge}
\end{equation}
\begin{equation}
\begin{gathered}
	\dot\rho_{e, \tilde{n}_{\alpha}, g, n}=-i\omega_{\sigma}\rho_{e, \tilde{n}_{\alpha}, g, n}+i\omega_{{\rm{v}}}(n-\tilde{n}_{\alpha})\rho_{e, \tilde{n}_{\alpha}, g, n}\\
+i\Omega e^{-i\omega t}\sum\limits_{n_{1}} \left\langle \tilde{n}_{\alpha}|n_{1}\right\rangle\rho_{g, n_{1}, g, n}\\
-i\Omega e^{-i\omega t}\sum\limits_{\tilde{n}_{2\alpha}} \left\langle \tilde{n}_{2\alpha}|n\right\rangle\rho_{e, \tilde{n}_{\alpha}, e, \tilde{n}_{2\alpha}}\\
-\frac{\gamma_{p}+\gamma_{D}+\gamma_{deph}}{2}\rho_{e, \tilde{n}_{\alpha}, g, n}
\end{gathered}
\label{eq:dot_rho_eg}
\end{equation}
\begin{equation}
\begin{gathered}
	\dot\rho_{g, n_{1}, g, n_{2}}=i\omega_{{\rm{v}}}(n_{1}-n_{2})\rho_{g, n_{1}, g, n_{2}}\\
+i\Omega e^{i\omega t}\sum\limits_{\tilde{n}_{1\alpha}} \left\langle n_{1}|\tilde{n}_{1\alpha}\right\rangle\rho_{e, \tilde{n}_{1\alpha}, g, n_{2}}\\
-i\Omega e^{-i\omega t}\sum\limits_{\tilde{n}_{2\alpha}} \left\langle \tilde{n}_{2\alpha}|n_{2}\right\rangle\rho_{g, n_{1}, e, \tilde{n}_{2\alpha}}\\
-\gamma_{p}\rho_{g, n_{1}, g, n_{2}}
+ \gamma_{D}\sum\limits_{\tilde{n}_{1\alpha}, \tilde{n}_{2\alpha}} \left\langle n_{1}|\tilde{n}_{1\alpha}\right\rangle \left\langle n_{2}|\tilde{n}_{2\alpha}\right\rangle \rho_{e, \tilde{n}_{1\alpha}, e, \tilde{n}_{2\alpha}}\\
	+\frac{\gamma_{{\rm{v}}}(\tilde{n}_{0}+1)}{2}\left(2\sqrt{(n_{1}+1)(n_{2}+1)}\rho_{g, n_{1}+1, g, n_{2}+1}\right.\\
-(n_{1}+n_{2})\rho_{g, n_{1}, g, n_{2}}\Bigr)\\
	+\frac{\gamma_{{\rm{v}}}\tilde{n}_{0}}{2}\left(2\sqrt{n_{1}n_{2}}\rho_{g, n_{1}-1, g, n_{2}-1}\right.\\
-(n_{1}+n_{2}+2)\rho_{g, n_{1}, g, n_{2}}\Bigr)
\end{gathered}
\label{eq:dot_rho_gg}
\end{equation}

\section{The dynamics of molecular polarization}\label{sec:results}
\subsection{Continuous-wave pumping}
First, we consider the case when only coherent external field excite the system, and there is no incoherent pumping, i.e., $\gamma_p = 0$.
We find the dynamics of the system by numerical simulations of the master equation~(\ref{eq:Lindblad_equation_first}).
As an initial condition, we consider the ground state of electronic subsystems and thermal state of vibron subsystem, $\hat \rho \left(0\right) = |g\rangle \langle g | \otimes \left(1-e^{-\hbar\omega_{{\rm{v}}}/kT}\right)\sum_n e^{-\hbar\omega_{{\rm{v}}}/kT} |n\rangle \langle n|$.
In addition, we consider the case of large detuning, $|\omega_{\sigma} - \omega| \ll \omega_{\sigma}, \omega$, which is typical for the nonresonant Raman scattering.
In the case when there is no interaction between vibrons and exciton ($g=0$), Eq.~(\ref{eq:Lindblad_equation_first}) describes the dynamics of exciton driven by monocromatic field.
\begin{figure}[H]
	\centering
	\begin{subfigure}{0.49\linewidth}
		\includegraphics[width=1\linewidth]{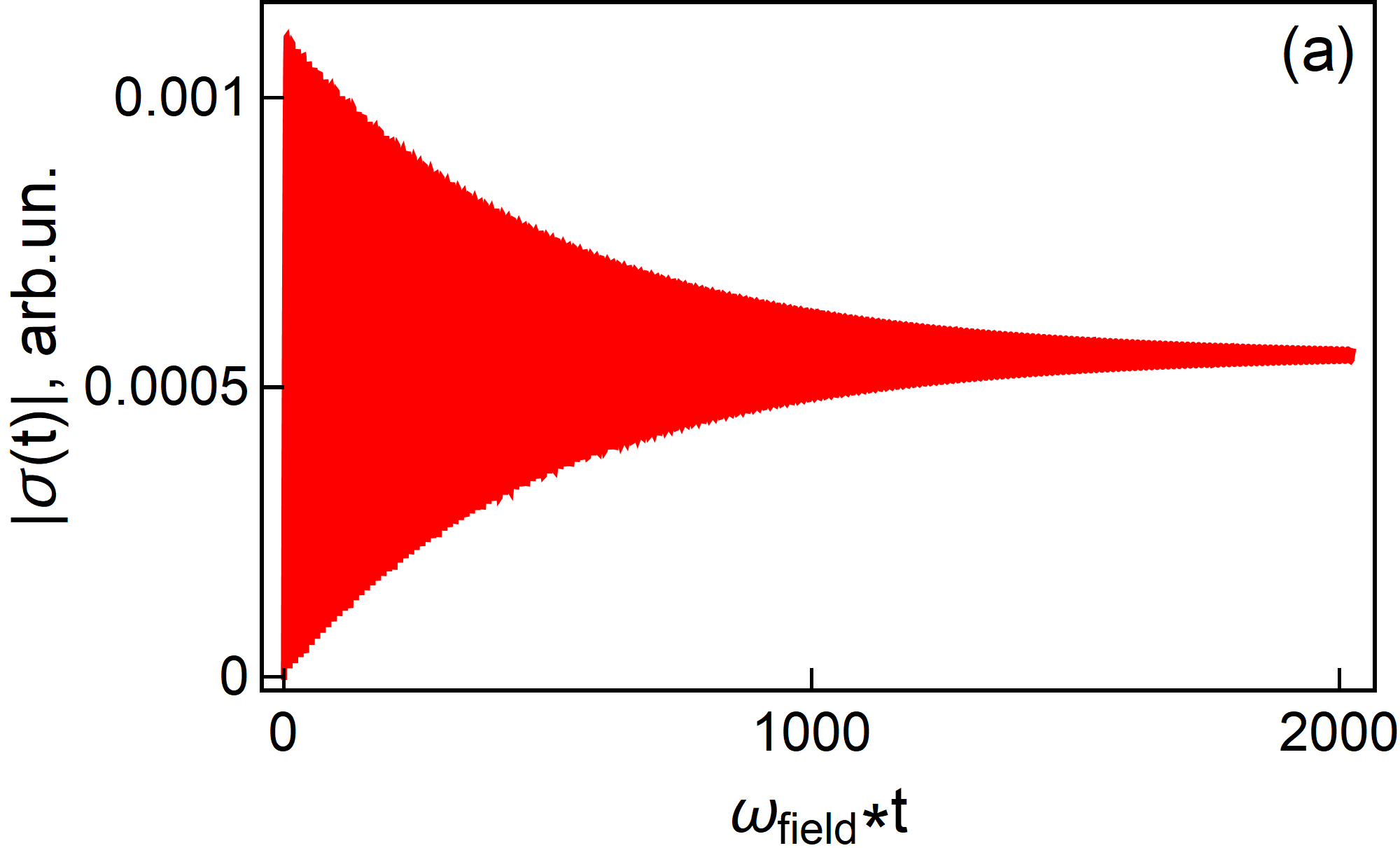}
		\phantomcaption
		\label{Dyng0}
	\end{subfigure}
	\hfill
	\begin{subfigure}{0.49\linewidth}
		\includegraphics[width=1\linewidth]{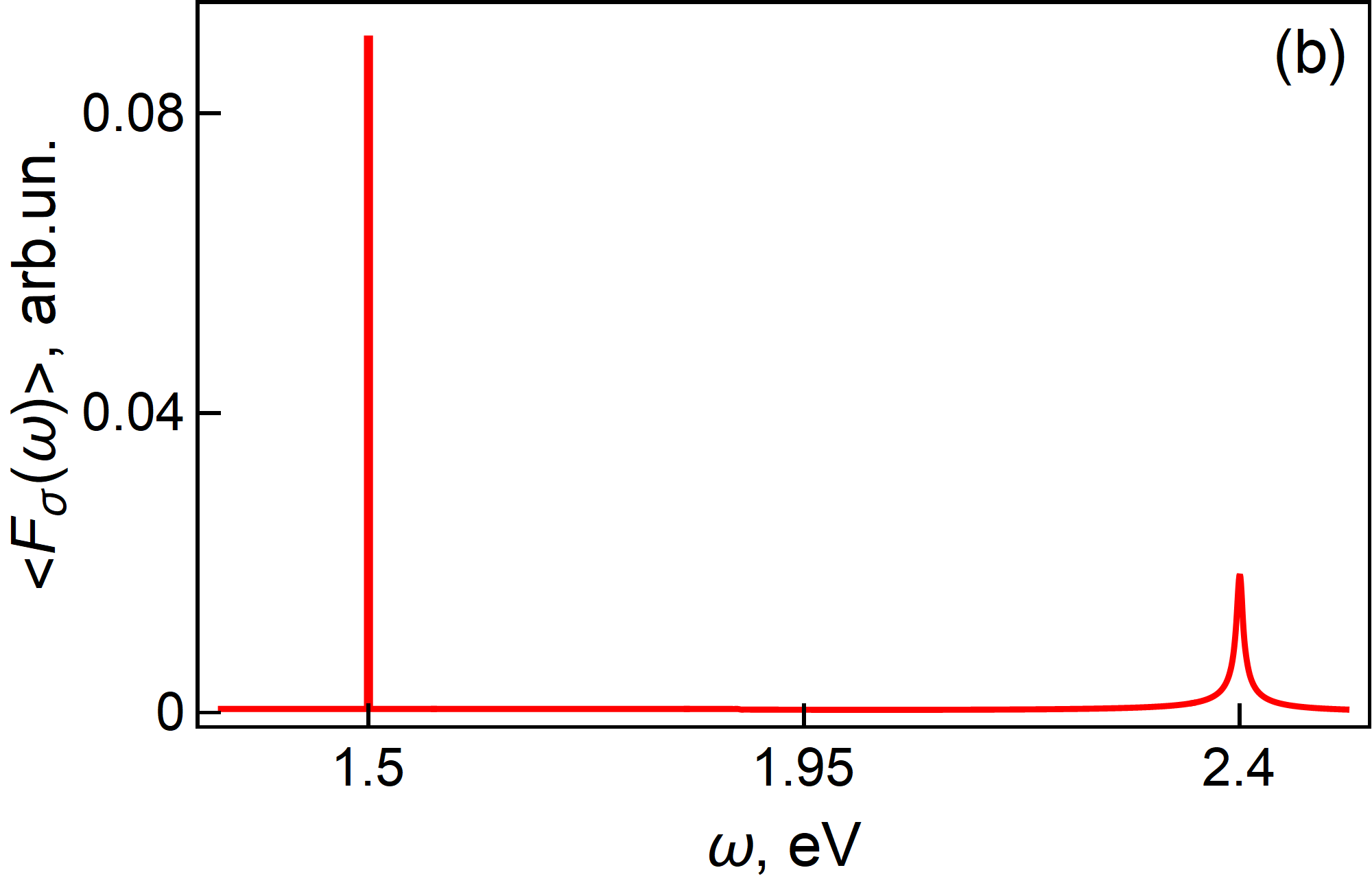}
		\phantomcaption
		\label{Dyng0Sp}
	\end{subfigure}	
	\caption{(a) The temporal dependence of $|\sigma\left(t\right)| = |{\rm{Tr}}\left(\hat \rho\left(t\right) \hat \sigma\right)|$, $\gamma_{D}=1\cdot10^{-3}~\text{eV}$, $\gamma_{deph}= 5\cdot10^{-3}~\text{eV}$, $\gamma_{{\rm{v}}}=2\cdot 10^{-4}~\text{eV}$, $\Omega_{R}=10^{-3}~\text{eV}$, $\omega_{field}=1.5~\text{eV}$, $\omega_{\sigma}=2.4~\text{eV}$, $\omega_{{\rm{v}}}=0.025~\text{eV}$, $g=0~\text{eV}$ (b) Spectrum of oscillations of $\sigma\left(t\right)$ }
\end{figure}
In such a case, the dynamics of molecular polarization, $\sigma\left(t\right) = {\rm{Tr}} \left(\hat \rho \left(t\right) \hat \sigma \right)$, can be divided~\cite{scully1997quantum} into damped Rabi oscillations at the exciton eigenfrequency and persistent driven oscillations at the frequency of the external field (Rayleigh response).
This dynamics are presented in Fig.~\ref{Dyng0}.
The spectrum of the oscillations is the sum of Lorentzian line at the exciton eigenfrequency with width equal to exciton transverse relaxation rates, and $\delta$-function corresponding to Rayleigh response, see Fig.~\ref{Dyng0Sp}.
The number of phonons remains unchanged, and there is no displacement of nuclei ($b=0$).

Let us now begin to increase the coupling constant between vibrons and excitons $g$.
\begin{figure}[H]
	\centering
	\begin{subfigure}{0.49\linewidth}
		\includegraphics[width=1\linewidth]{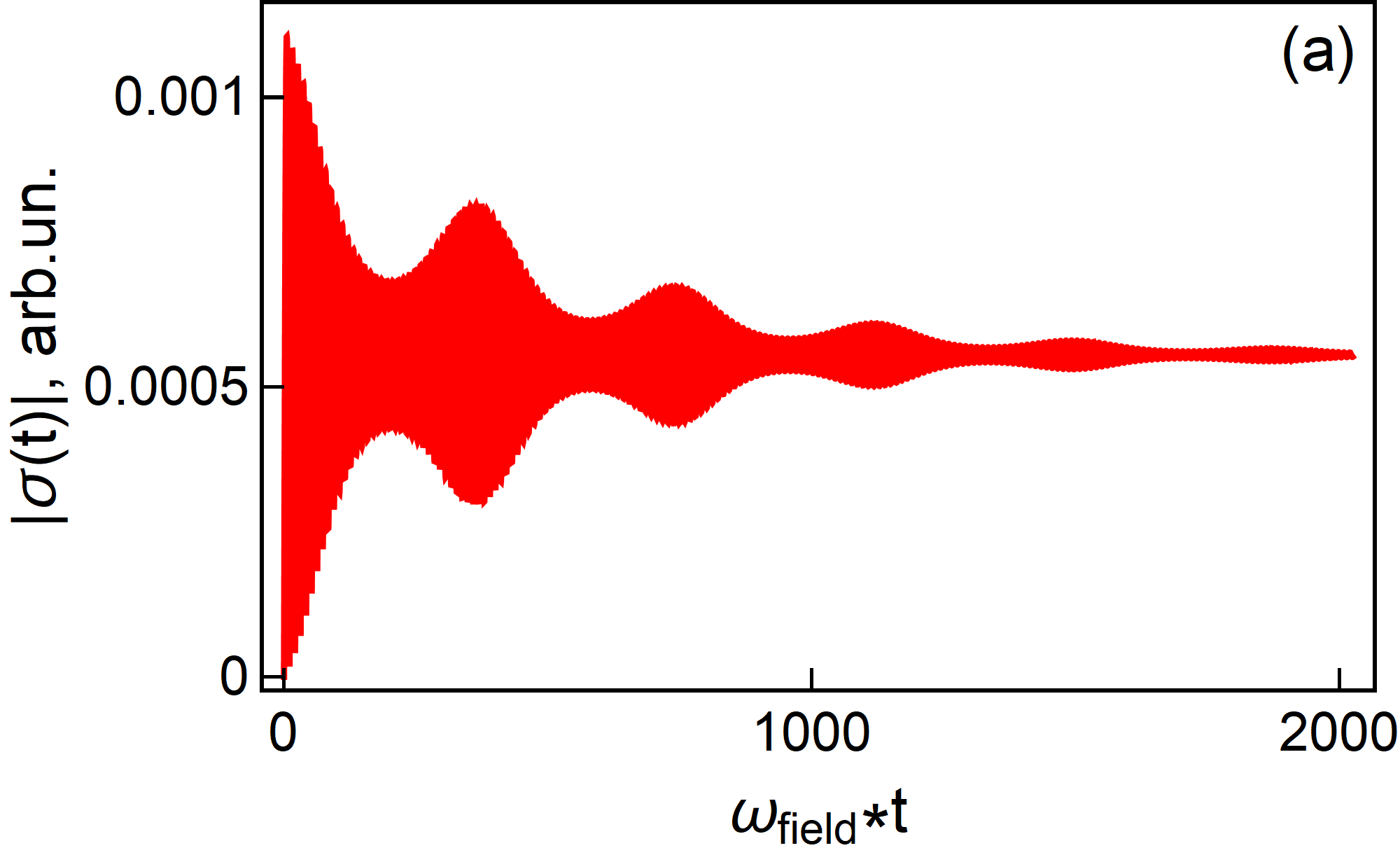}
		\phantomcaption
		\label{Dyng1}
	\end{subfigure}
	\hfill
	\begin{subfigure}{0.49\linewidth}
		\includegraphics[width=1\linewidth]{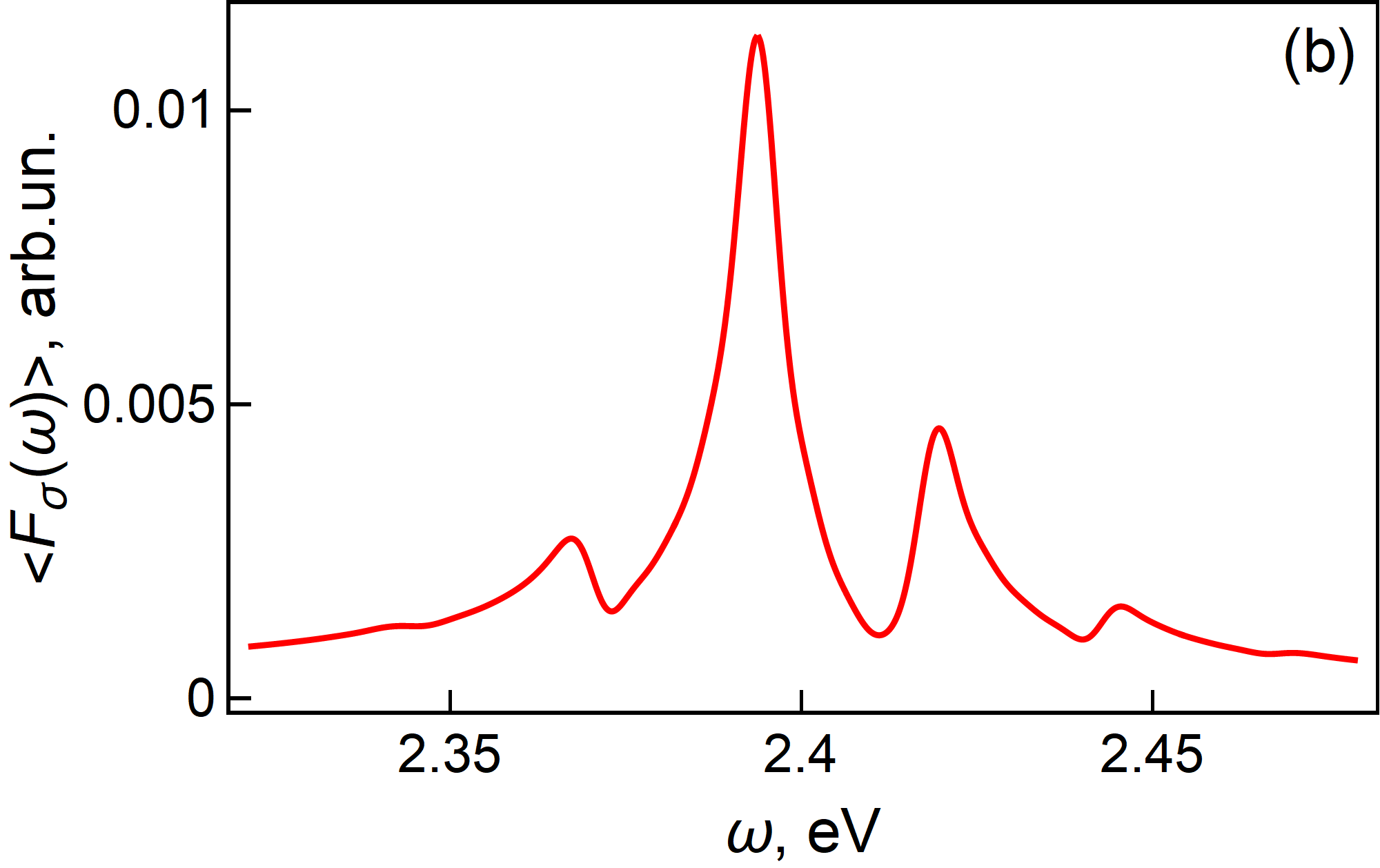}
		\phantomcaption
		\label{Dyng1Sp}
	\end{subfigure}	
	\caption{(a) The temporal dependence of $|\sigma\left(t\right)| = |{\rm{Tr}}\left(\hat \rho\left(t\right) \hat \sigma\right)|$, $\gamma_{D}=1\cdot10^{-3}~\text{eV}$, $\gamma_{deph}= 5\cdot10^{-3}~\text{eV}$, $\gamma_{{\rm{v}}}=2\cdot 10^{-4}~\text{eV}$, $\Omega_{R}=10^{-3}~\text{eV}$, $\omega_{field}=1.5~\text{eV}$, $\omega_{\sigma}=2.4~\text{eV}$, $\omega_{{\rm{v}}}=0.025~\text{eV}$, $g=0.0125~\text{eV}$ (b) Spectrum of oscillations of $|\sigma\left(t\right)|$ }
\end{figure}
At small values of $g \ll \omega_{{\rm{v}}}$, dipole moment also exhibits Rabi oscillations but with beating at vibron frequency, see Fig.~\ref{Dyng1}.
The reason of this beating is that the electric field by itself results in oscillations between electronic ground and excited states.
Because direct product of Fock vibron state and excited electronic state is not the molecular eigenstate, the system begins to oscillate at the eigenfrequencies described by Eq.~(\ref{ShiftStatesEx}).
The difference between these eigenfrequencies is vibron eigenfrequency $\omega_{{\rm{v}}}$.
The amplitude of vibrons oscillates at vibron eigenfrequency $\omega_{{\rm{v}}}$.
In the spectrum, the multiple peaks near the exciton transition frequency $\omega_\sigma$ appear, see Fig.~\ref{Dyng1Sp} (in this and subsequent figures, we do not show the $\delta$-function of Rayleigh response).

Further increasing of the coupling constant between phonons and the dipole moment $g$ results in intensifying of beating of the dipole moment.

\begin{figure}[H]
	\centering
	\begin{subfigure}{0.49\linewidth}
		\includegraphics[width=1\linewidth]{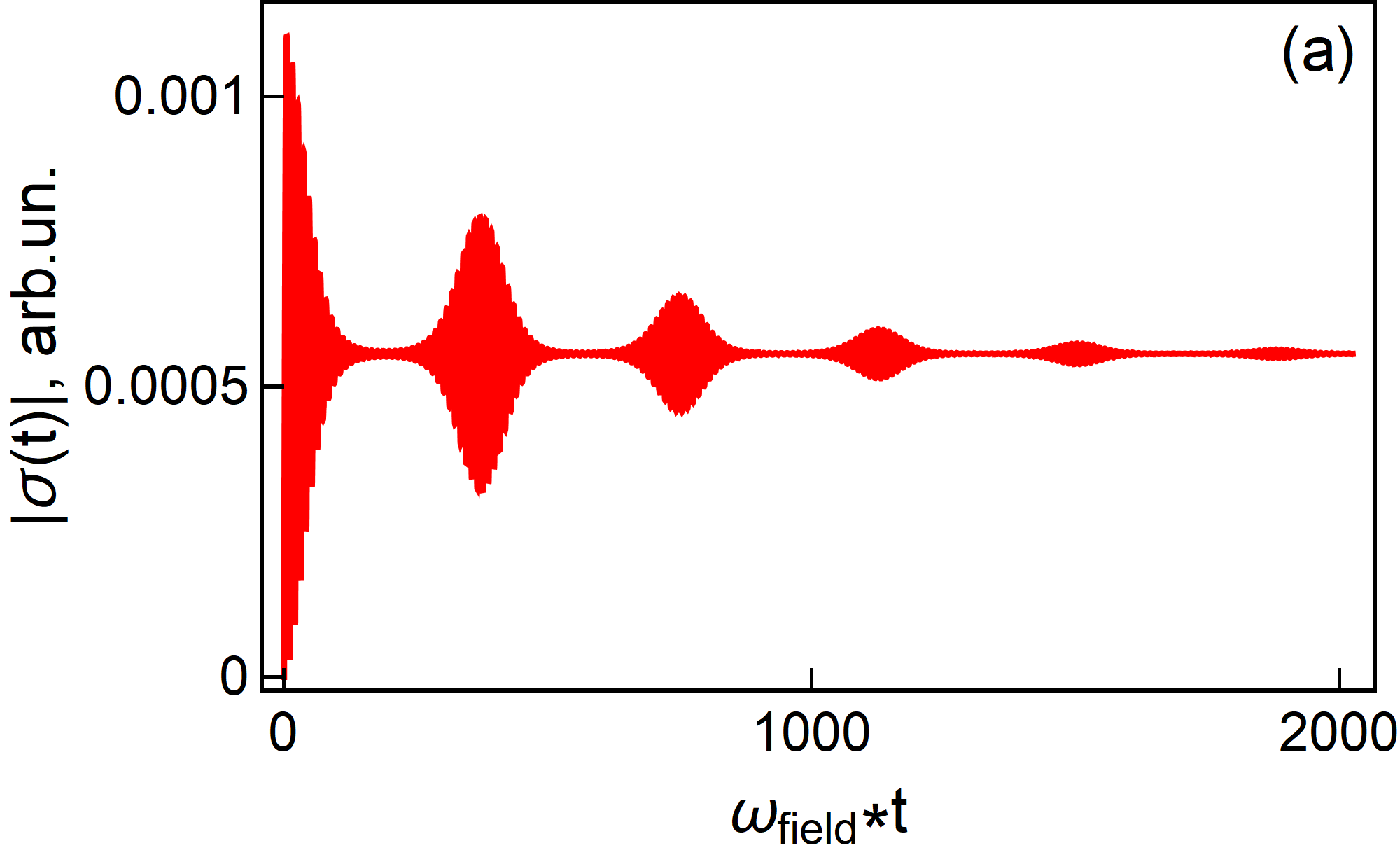}
		\phantomcaption
		\label{Dyng2}
	\end{subfigure}
	\hfill
	\begin{subfigure}{0.49\linewidth}
		\includegraphics[width=1\linewidth]{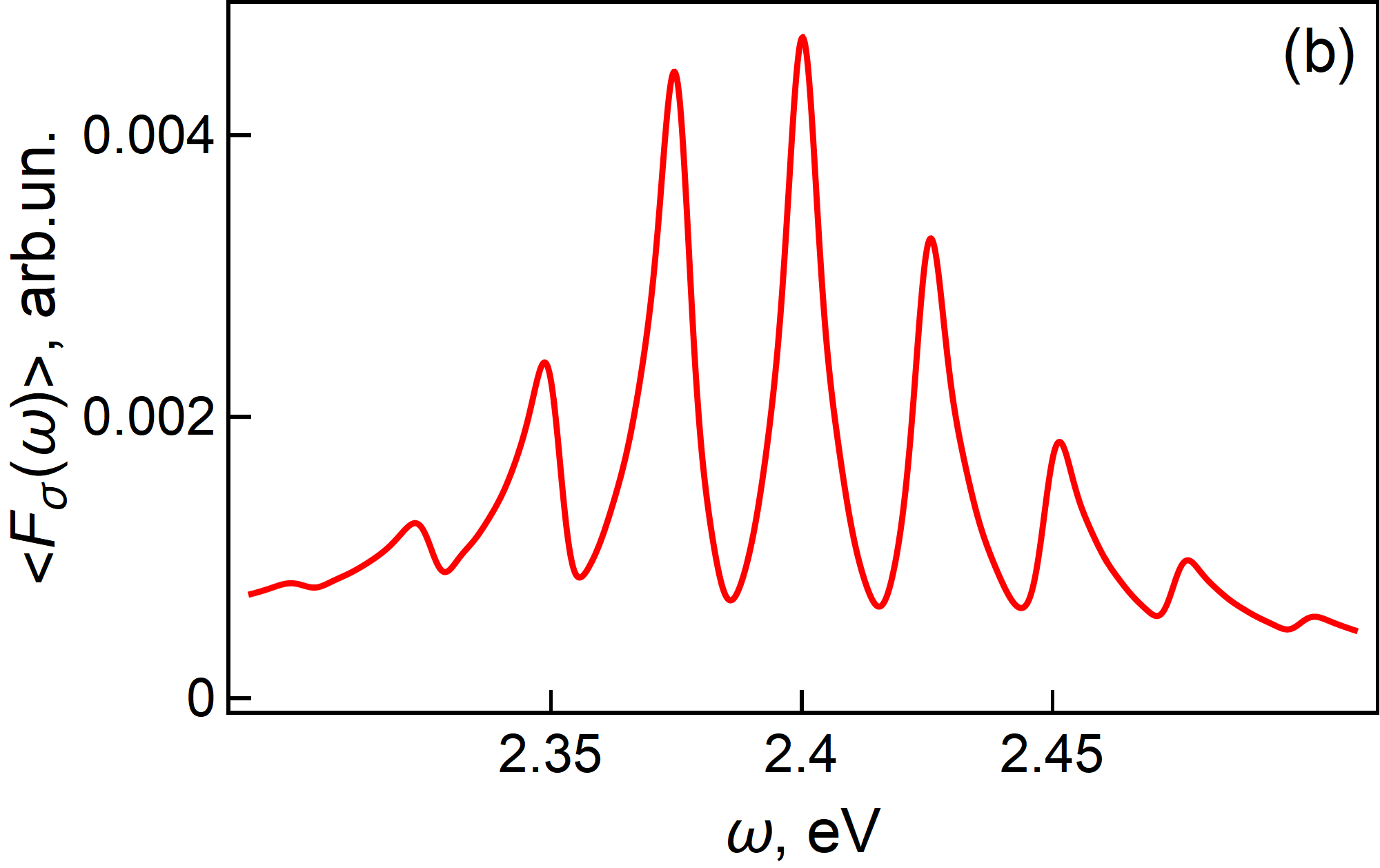}
		\phantomcaption
		\label{Dyng2Sp}
	\end{subfigure}	
	\caption{(a) The temporal dependence of $|\sigma\left(t\right)| = |{\rm{Tr}}\left(\hat \rho\left(t\right) \hat \sigma\right)|$, $\gamma_{D}=1\cdot10^{-3}~\text{eV}$, $\gamma_{deph}= 5\cdot10^{-3}~\text{eV}$, $\gamma_{{\rm{v}}}=2\cdot 10^{-4}~\text{eV}$, $\Omega_{R}=10^{-3}~\text{eV}$, $\omega_{field}=1.5~\text{eV}$, $\omega_{\sigma}=2.4~\text{eV}$, $\omega_{{\rm{v}}}=0.025~\text{eV}$, $g=0.025~\text{eV}$ (b) Spectrum of oscillations of $|\sigma\left(t\right)|$ }
\end{figure}

When the coupling constant between exciton and vibron $g$ becomes larger than vibron eigenfrequency $\omega_{{\rm{v}}}$, beating of exciton dipole moment transforms to repeating collapses and revivals, see Fig.~\ref{Dyng2}.
The spectrum of the system is the sum of the Rayleigh $\delta$-peak and multiple peaks at the exciton transition frequency, see Fig.~\ref{Dyng2Sp}. 

In all of these cases, beating of oscillations and collapses and revivals dissipate with the exciton dephasing rate.
We verify this by changing dephasing rate and observing that beating and collapses and revivals relax with this rate.

\begin{figure}[H]
	\centering
	\begin{subfigure}{0.49\linewidth}
		\includegraphics[width=1\linewidth]{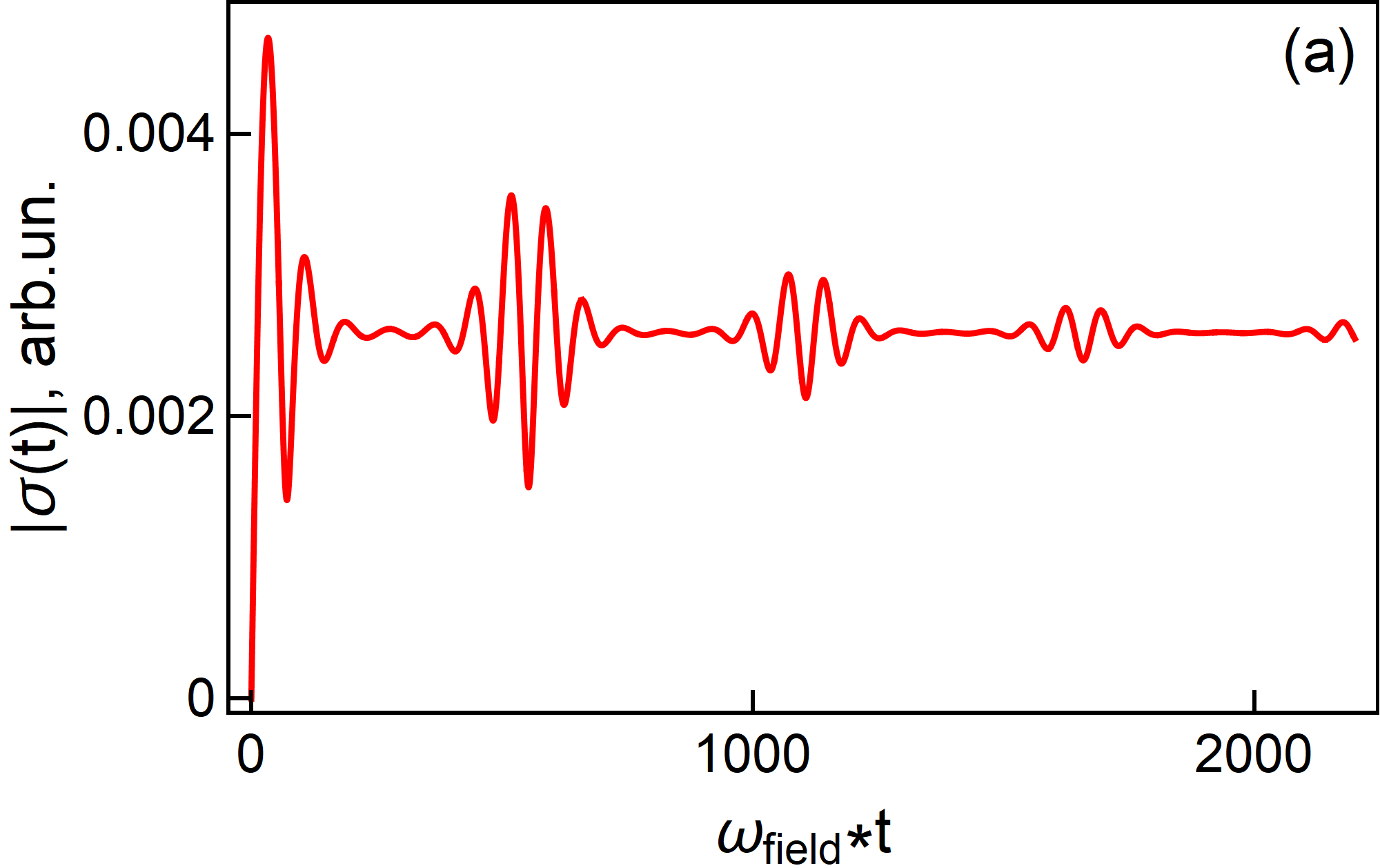}
		\phantomcaption
		\label{Dyng2D}
	\end{subfigure}
	\hfill
	\begin{subfigure}{0.49\linewidth}
		\includegraphics[width=1\linewidth]{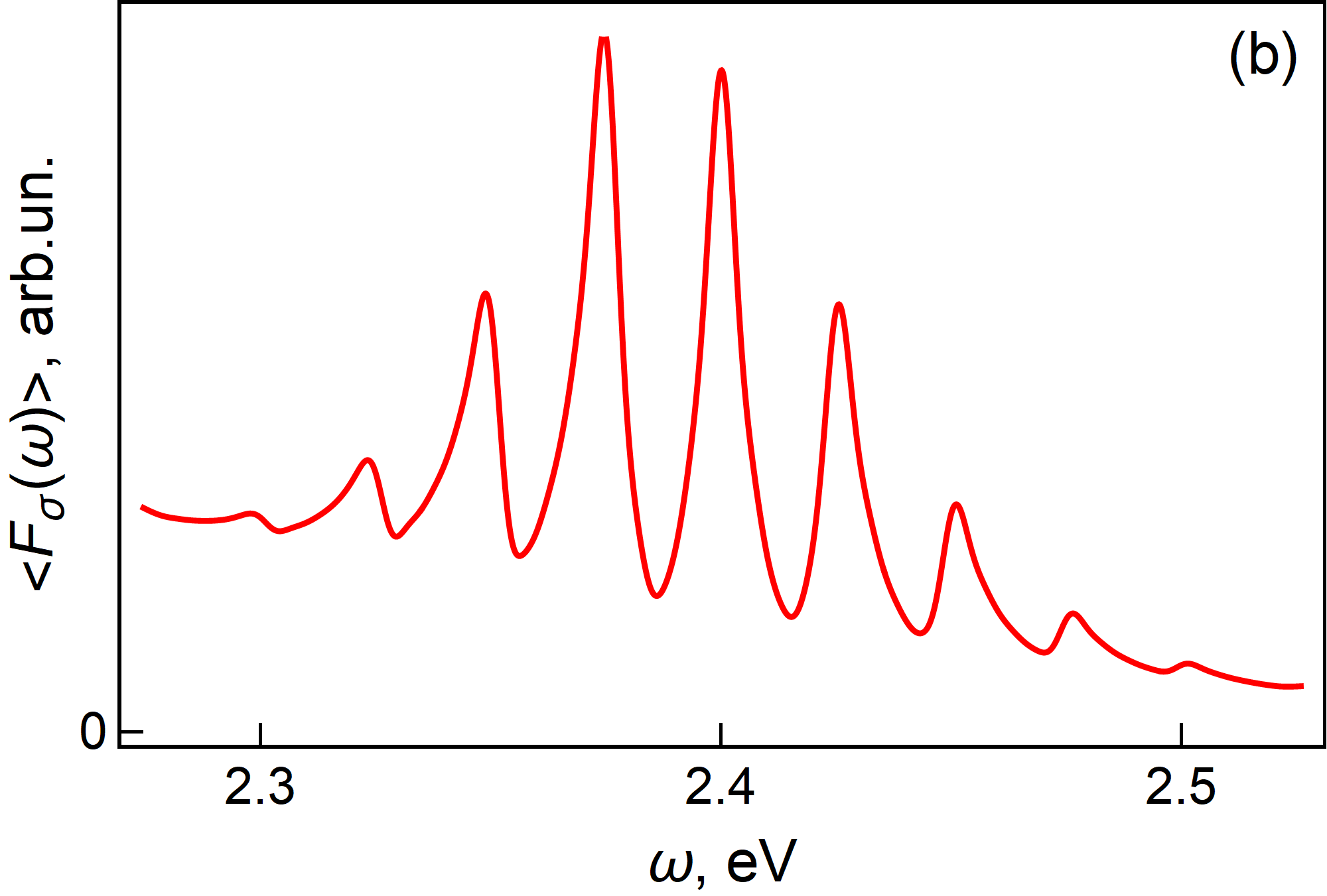}
		\phantomcaption
		\label{Dyng2DSp}
	\end{subfigure}	
	\caption{(a) The temporal dependence of $|\sigma\left(t\right)| = |{\rm{Tr}}\left(\hat \rho\left(t\right) \hat \sigma\right)|$, $\gamma_{D}=1\cdot10^{-3}~\text{eV}$, $\gamma_{deph}= 5\cdot10^{-3}~\text{eV}$, $\gamma_{{\rm{v}}}=2\cdot 10^{-4}~\text{eV}$, $\Omega_{R}=10^{-3}~\text{eV}$, $\omega_{field}=2.2~\text{eV}$, $\omega_{\sigma}=2.4~\text{eV}$, $\omega_{{\rm{v}}}=0.025~\text{eV}$, $g=0.025~\text{eV}$ (b) Spectrum of oscillations of $|\sigma\left(t\right)|$ }
\end{figure}

Note that when the frequency of the incident field $\omega$ approaches to the exciton frequency, $\omega_{\sigma}$, the behavior qualitatively remains the same, see Fig.~\ref{Dyng2D}, but the amplitude of multiple peaks near $\omega_{\sigma}$ becomes smaller, see Fig.~\ref{Dyng2DSp}.

\subsection{Pulse pumping}
Let us now consider the case when in addition to the coherent pump there is short pump pulse.
We suppose that the rate of incoherent pumping is
\begin{gather}
	\gamma_{p}= \gamma^0_{p}, 0 \leq t \leq t_0;\quad \gamma_p = 0, t \geq t_0
\end{gather}

\begin{figure}[H]
	\centering
	\begin{subfigure}{0.32\linewidth}
		\includegraphics[width=1\linewidth]{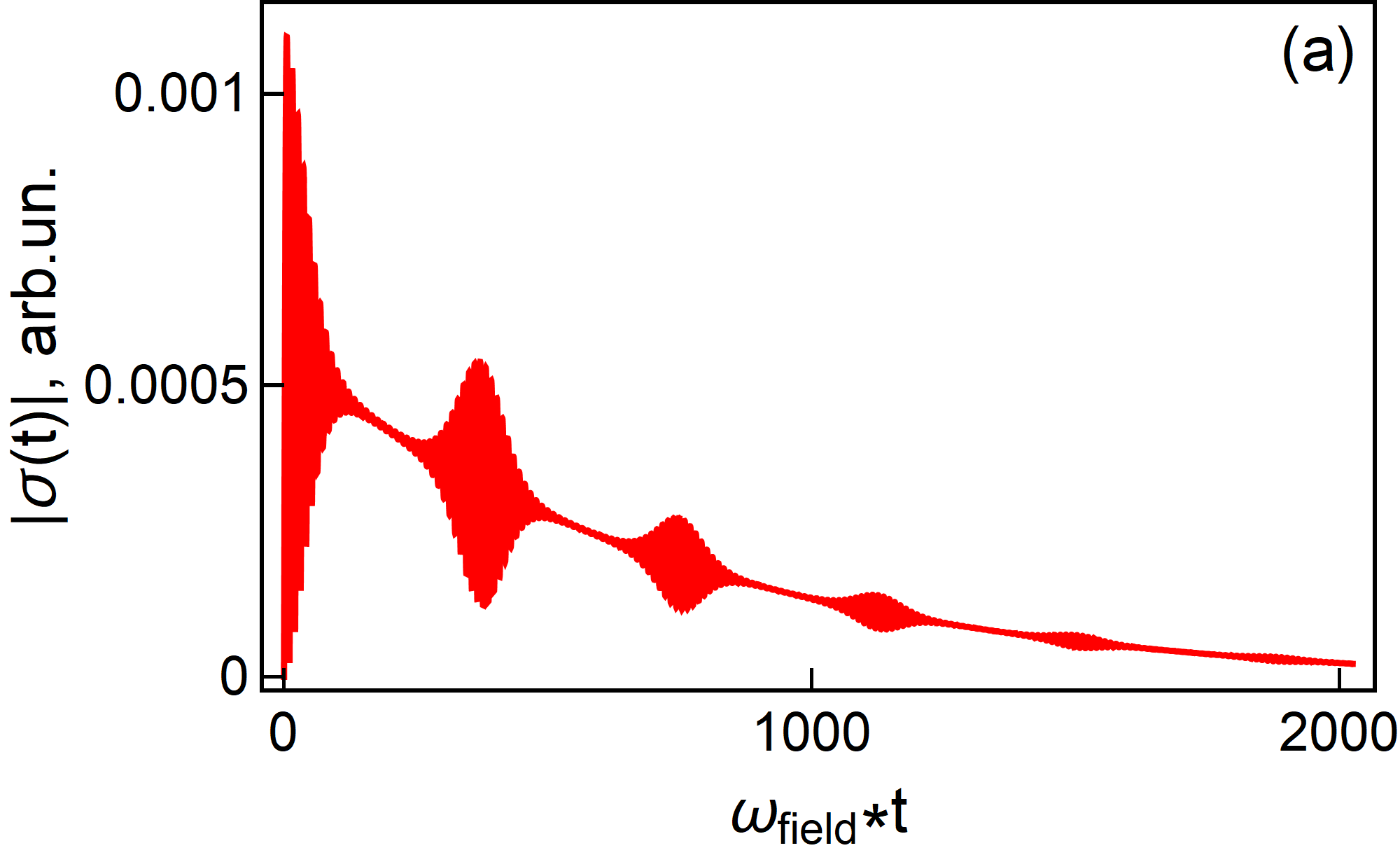}
		\phantomcaption
		\label{Dyng2Incoh}
	\end{subfigure}
	\hfill
	\begin{subfigure}{0.32\linewidth}
		\includegraphics[width=1\linewidth]{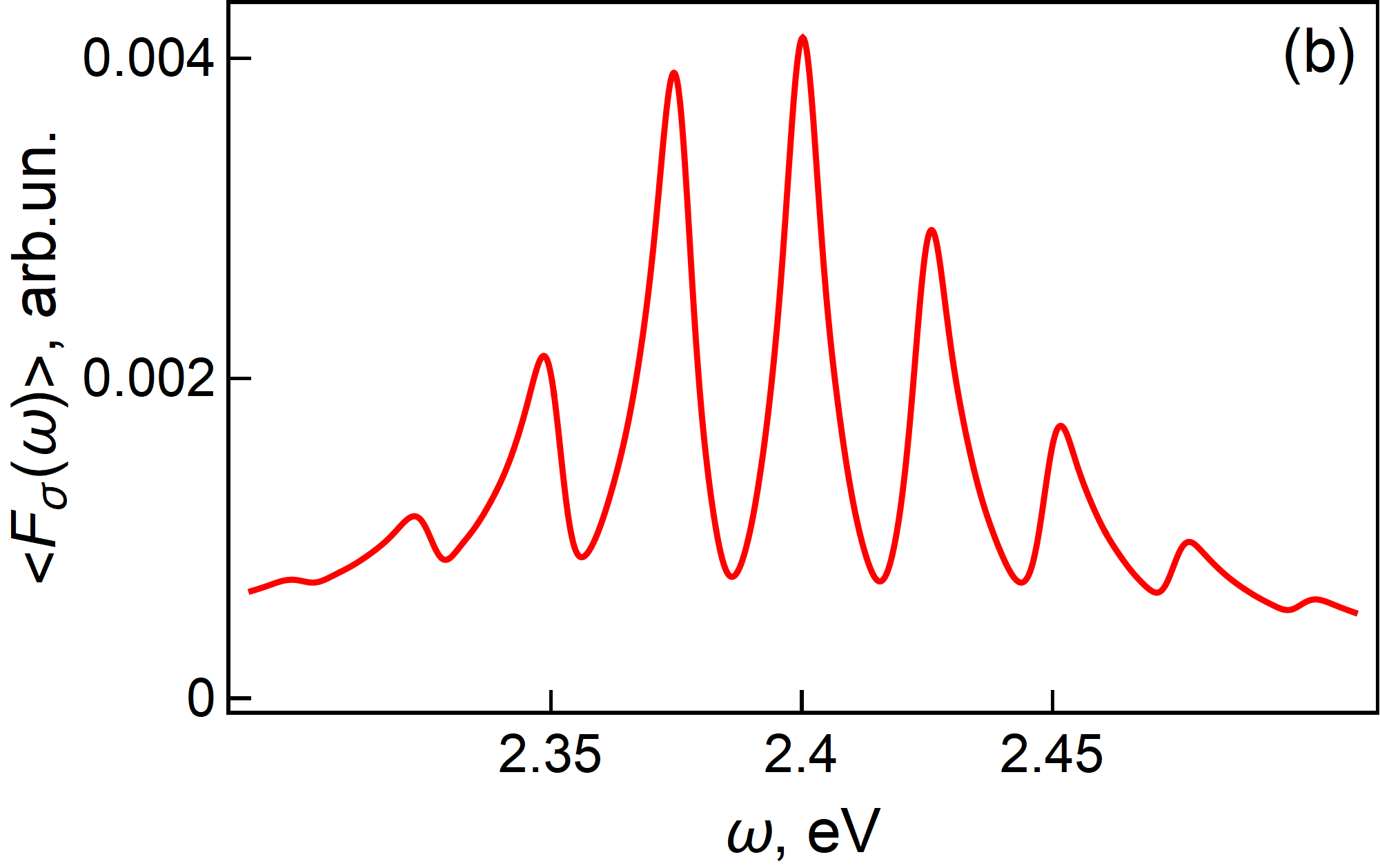}
		\phantomcaption
		\label{Dyng2IncohSp}
	\end{subfigure}	
	\hfill
	\begin{subfigure}{0.32\linewidth}
		\includegraphics[width=1\linewidth]{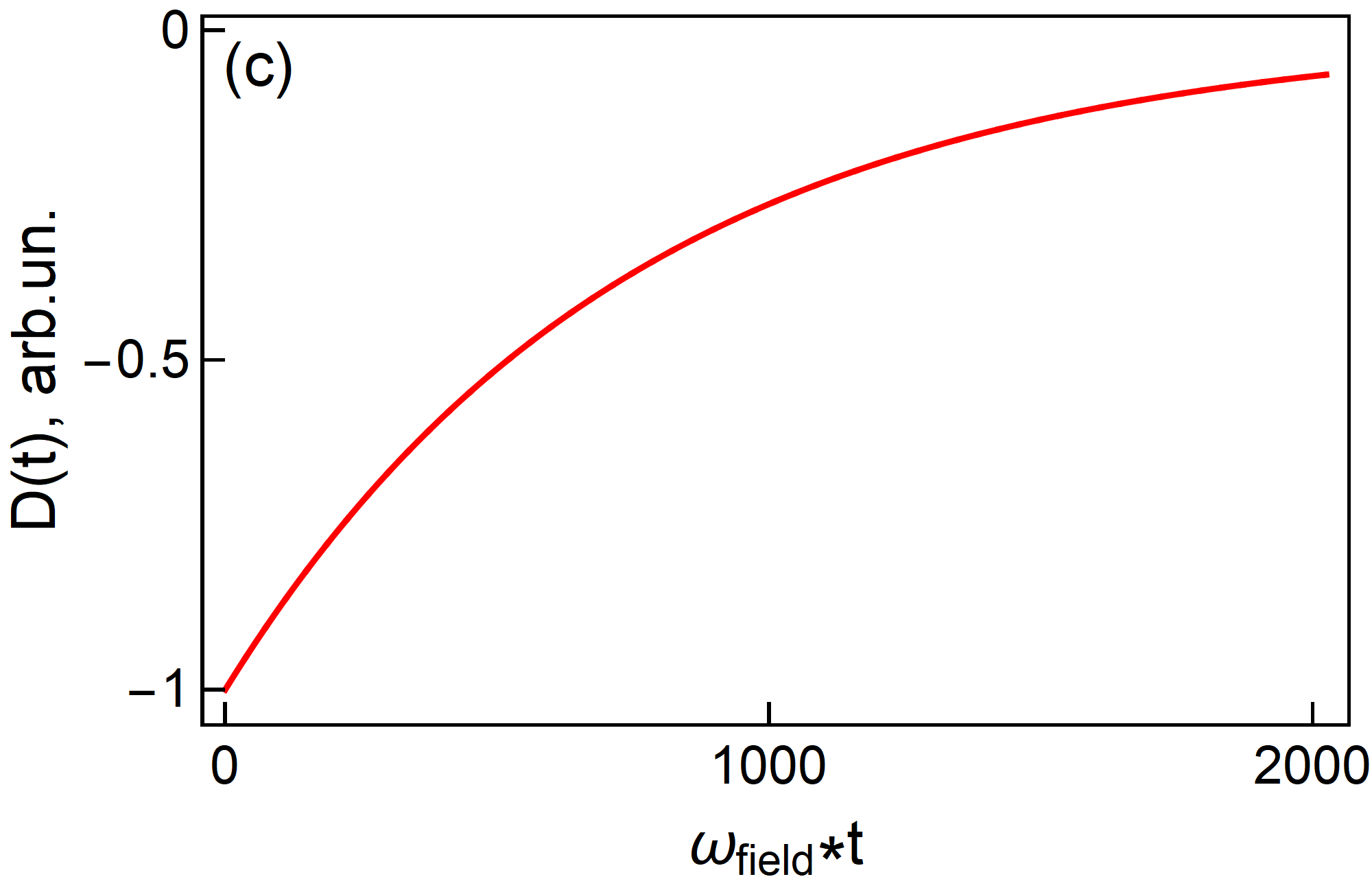}
		\phantomcaption
		\label{Dyng2IncohD}
	\end{subfigure}
	\caption{(a) Temporal dynamics of $|\sigma\left(t\right)|$, $\gamma_{D}=10^{-3}~\text{eV}$, $\gamma_{deph}=5\cdot10^{-3}~\text{eV}$, $\gamma_{{\rm{v}}}=2\cdot10^{-4}~\text{eV}$, $\Omega_{R}=10^{-3}~\text{eV}$, $\omega_{field}=1.5~\text{eV}$, $\omega_{\sigma}=2.4~\text{eV}$, $\omega_{{\rm{v}}}=0.025~\text{eV}$, $g=0.025~\text{eV}$ (b) Spectrum of oscillation of $\sigma\left(t\right)$; (c) Temporal dynamics of $D\left(t\right) = {\rm{Tr}}\left(\hat \rho\left(t\right) \hat D\right)$.}
\end{figure}

The dynamics of the molecule polarization and its spectrum are shown in Fig.~\ref{Dyng2Incoh},\subref{Dyng2IncohSp}.
During the pump pulse, the population of the excited state increases which results in saturation of excited state population, see Fig.~\ref{Dyng2IncohD}.
After pump pulse is turned off, population inversion gets smaller and the polarization starts to grow.
During these processes, the polarization dynamics exhibit collapses and revivals similar to the case without of pulse pump.
The only difference is during saturation the polarization amplitude decreases and amplitude of oscillations decreases too. 

\section{Estimation of the collapse and revivals times}
To estimate collapse and revival times, we, first, demonstrate that it is Hermitian part of the system dynamics which is responsible for the collapses and revivals.
To do that, we write the Schrödinger equation for the wave function of excitons and vibrons, which has the form:
\begin{equation}\label{ShroedEq}
i\hbar\frac{\partial}{\partial t}\left|\psi\right>=\hat{\tilde{H}}\left|\psi\right>
\end{equation}
where we expand the wave function over the eigenstates of $\hat H_{{\rm{mol}}}$
\begin{equation}\label{ShroedEqExp}
\left|\psi\right>=\sum\limits_{n}{C_{g,n}(t)\left|g,n\right>} + \sum\limits_{\tilde{n}_{\alpha}}{C_{e,\tilde{n}_{\alpha}}(t)\left|e,\tilde{n}_{\alpha}\right>}
\end{equation}
The system Hamiltonian is determined by Eq.~(\ref{eq:Hamiltonian}).

\begin{figure}[H] 
	\includegraphics[width=1\linewidth]{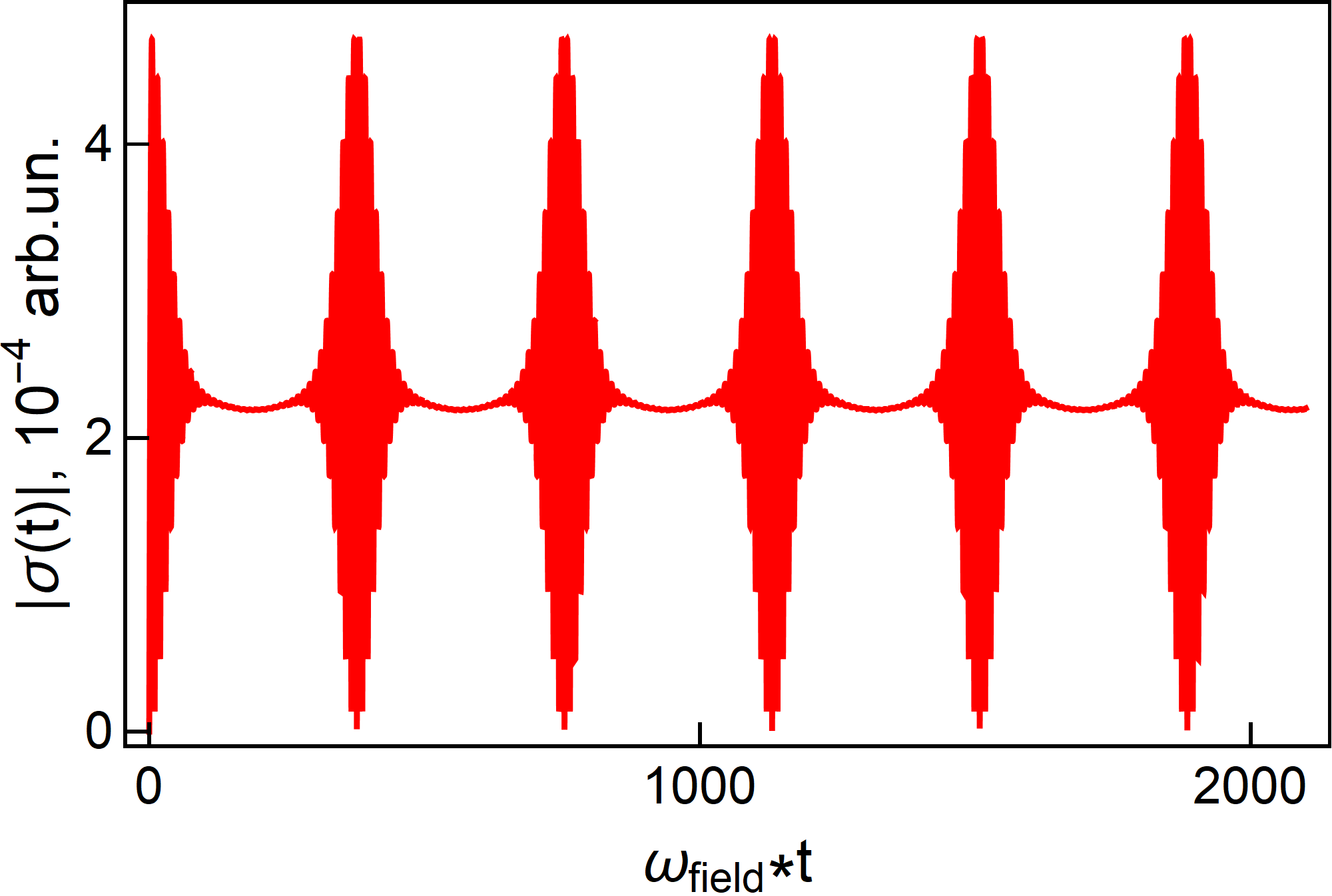}
	\caption{Temporal dynamics of $|\sigma\left(t\right)|$, $\Omega_{R}=10^{-3}~\text{eV}$, $\omega_{field}=1.5~\text{eV}$, $\omega_{\sigma}=2.4~\text{eV}$, $\omega_{{\rm{v}}}=0.025~\text{eV}$, $g=0.05~\text{eV}$}
	\label{HermDyn}
\end{figure}

The dynamics of mean value of $\hat \sigma$ obtained from numerical simulations of Eqs.~(\ref{ShroedEq})-(\ref{ShroedEqExp}) are shown in Fig.~\ref{HermDyn}.
It is seen that Hermitian dynamics quantitatively reproduce the dynamics obtained from the Lindblad equation~(\ref{eq:Lindblad_equation_first}) at times smaller than inverse dephasing rate $\gamma_{\sigma}$.
Namely, there is the collapses in the dynamics of molecule polarization and the revivals that occur at the same times as they occur when modeling the Lindblad equation~(\ref{eq:Lindblad_equation_first}).
Thus, we can conclude that it is Hemrmitian part of dynamics that is responsible for formation of collapses and revivals, and neglect dissipation processes to estimate collapse and revival times.

\begin{figure}[H]
	\includegraphics[width=1\linewidth]{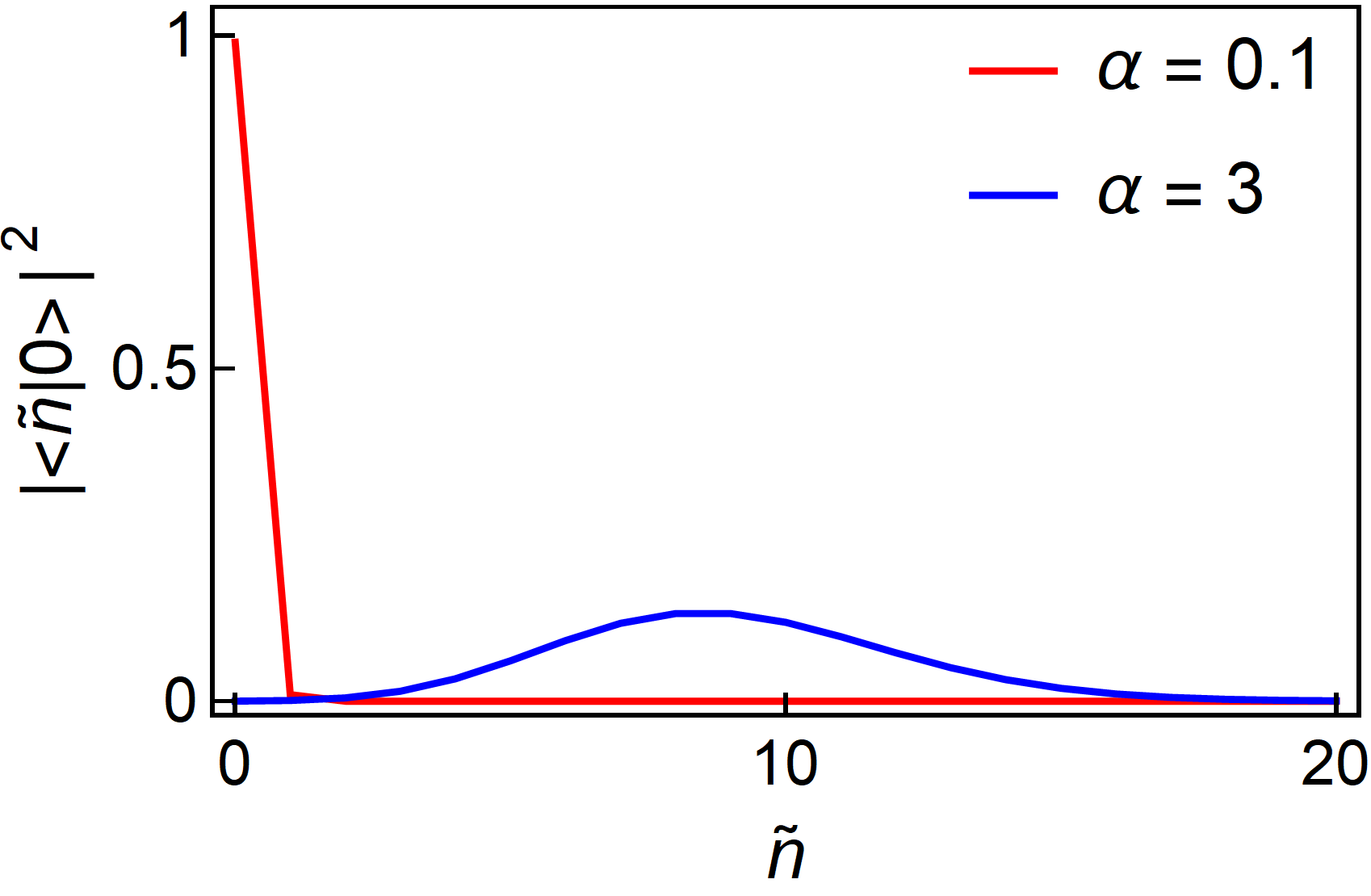}
	\caption{Dependence of the matrix element $\langle \tilde n | n \rangle$ on $\tilde n$ for $n = 0$}
\label{ShiftFockMatr}
\end{figure}

To estimate the collapse time, we use the following reasoning.
The eigenstates of the system are described by Eqs.~(\ref{ShiftStatesGround})-(\ref{ShiftStatesEx}).
Consider the case when the system is initially in the state $|g,0\rangle$.
The electric field results in transitions to the state $|e, 0\rangle$.
This state is not an eigenstate of the system.
As a result, the projection of the system state to the eigenstate, say, $|e,\tilde n_{\alpha} \rangle$, will oscillate with the frequency of this eigenstate and will have the amplitude $\langle \tilde n_{\alpha} | 0 \rangle$.
In Fig.~\ref{ShiftFockMatr}, we show the dependence of this matrix element in two cases: weak, $g/\omega_{{\rm{v}}} \ll 1$ and strong $g/\omega_{{\rm{v}}} \gg 1$ exciton-vibron coupling.
We see that the higher the value $g/\omega_{{\rm{v}}}$, the smoother the function $\langle \tilde n_{\alpha} | 0 \rangle$.
Thus, in the case of strong exciton-vibron coupling, after the transition to the state $|e,0\rangle$, one has simultaneous excitations of many system eigenstates $|e,\tilde n_{\alpha} \rangle$.
The projections of the system state to these eigenstates will oscillate with close eigenfrequencies $\omega_{\sigma}\left(1-\alpha^2\right) + \omega_{{\rm{v}}}\left(\tilde n_{\alpha} + 1/2\right)$.
The difference between these eigenfrequencies, $\omega_{{\rm{v}}}$ is much less than eigenfrequencies themselves.

Thus, we can consider the shifted Fock states as an effective reservoir for the state $\left|e,0\right>$ which external field excites.
Indeed, analogous situation takes place if we consider one initially excited harmonic oscillator interacting with an interaction constant $g$ with a set of large number of harmonic oscillators with difference $\Delta \omega$ between their eigenfrequencies.
The last will play the role of effective reservoir and will result in exponential relaxation of the oscillator energy with the rate $\pi g^2/\Delta \omega$~\cite{sergeev2021new}.

As a result, we can estimate collapse time as an inverse rate of relaxation of oscillations to the reservoir of system eigenstates:
\begin{gather}\label{ColTime}
	t_{{\rm{col}}} \simeq \frac{\omega_{{\rm{v}}}}{\pi g^2 \left| M\left(\alpha\right)\right|^2} \sim \left(g \alpha \left|M\left(\alpha\right)\right|^2\right)^{-1}
\end{gather}
where $M\left(\alpha\right) = {\rm{max}} \langle 0| \tilde n_{\alpha} \rangle$ is characteristic value of matrix elements of transition described by Eq.~(\ref{MatrEl}).

To estimate the revival time, we should take into account that the effective reservoir of shifted vibron states has discrete spectrum. 
It is known that in such a case, the excitation, initially stored in system and subsequently transferred into reservoir, will return to the system after the time $~2\pi/\Delta\omega$~\cite{tatarskiui1987example}.
Thus, in the considered case, the revival time can be estimated as $2\pi/\omega_{{\rm{v}}}$.

\begin{figure}[H]
	\centering
	\begin{subfigure}{0.49\linewidth}
		\includegraphics[width=1\linewidth]{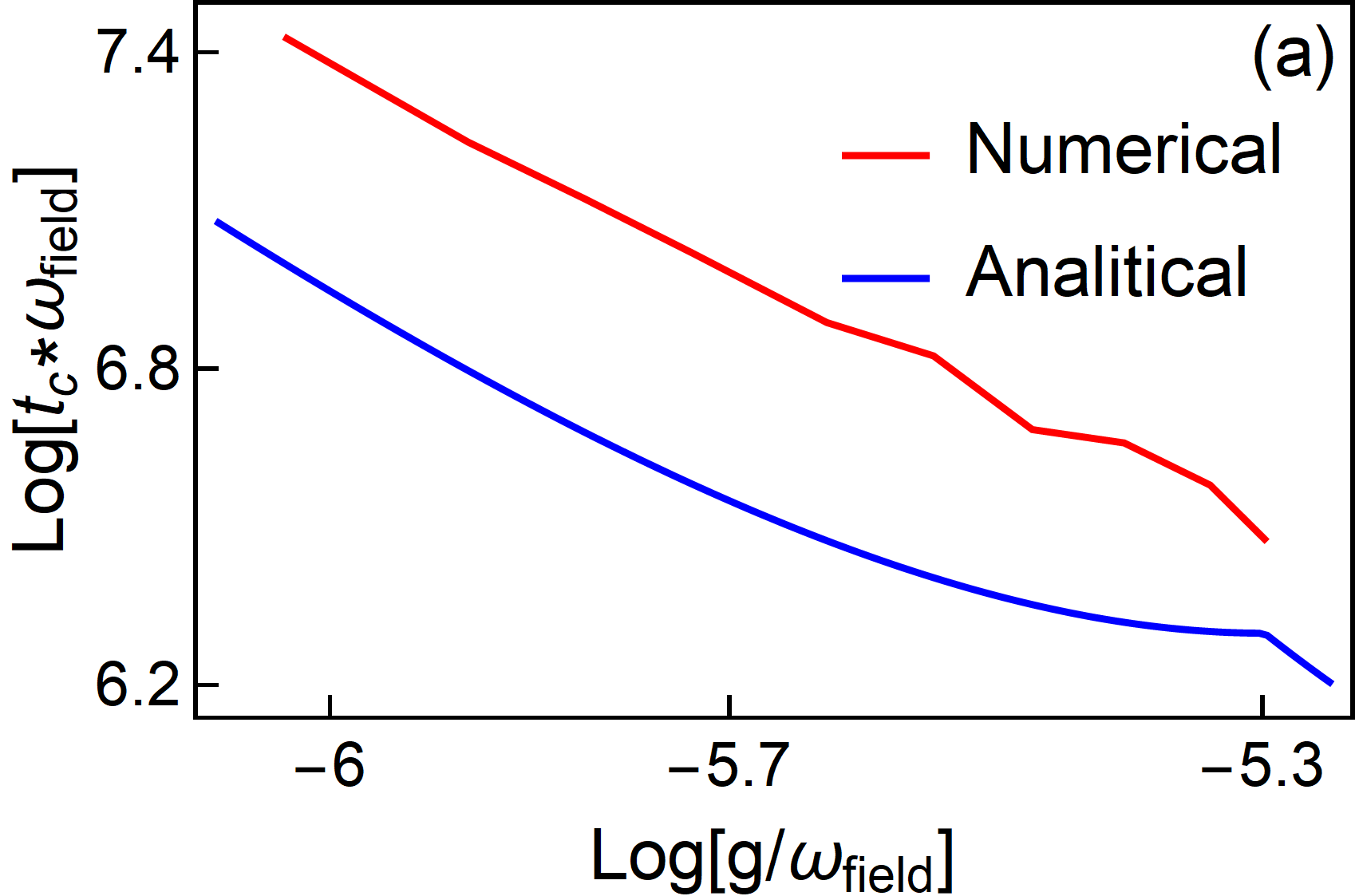}
		\phantomcaption
		\label{ColTimeg}
	\end{subfigure}
	\hfill
	\begin{subfigure}{0.49\linewidth}
		\includegraphics[width=1\linewidth]{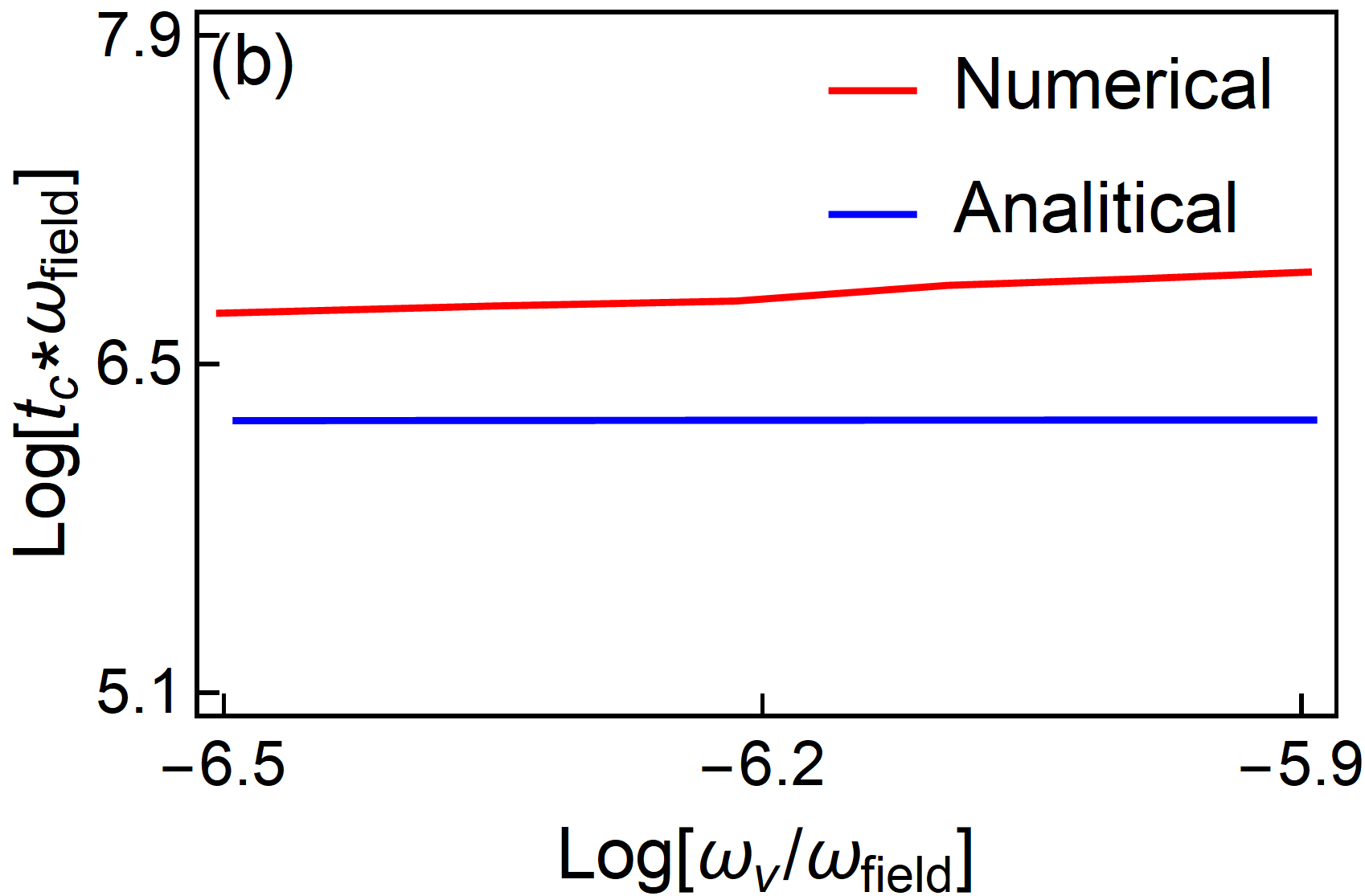}
		\phantomcaption
		\label{ColTimeOm}
	\end{subfigure}	
	\caption{ The theoretical dependence of the collapse time on the Frohlich constant $g$ (a) and on vibron eigenfrequency $\omega_{{\rm{v}}}$ in logarithmic scale, $\Omega_{R}=10^{-3}~\text{eV}$, $\omega_{field}=1.5~\text{eV}$, $\omega_{\sigma}=2.4~\text{eV}$, $\omega_{{\rm{v}}}=0.025~\text{eV}$, $g=0.025~\text{eV}$}
\end{figure}

In Fig.~\ref{ColTimeg},\subref{ColTimeOm} it is shown the dependence of the time of the first collapse and revival as a function of Frohlich constant of interaction and vibron eigenfrequency obtained from numerical simulation and from analytical estimations.
It is seen good agreement between them.


\section{CONCLUSION}
In this work, we consider the temporal dynamics of the molecule which electronic degrees of freedom strongly coupled with degrees of freedom of nuclei vibrations.
We show that when the coupling constant is larger than the vibron frequency, transient oscillations of polarization exhibit collapses.
The reason for the collapses is connected with structure of the eigenstates of the system.
The eigenstates of the system can be divided into two subsets.
The first one is the direct product of the ground electronic state and Fock states of nuclei vibrations.
The second subset is the direct product of the excited electronic state and shifted Fock states of nuclei vibrations.
The action of external monochromatic field on the molecular dipole moment by itself results in Rabi cycle between ground and excited electronic states during which the system change the electronic state, but does not change vibrational state because we consider Raman active molecules.
As a result, after half of a Rabi cycle, starting from, e.g., ground electronic state and Fock vibrational state, the system occurs at excited electronic state and still Fock vibrational state.
This state is not an eigenstate and the system and the system begins to transit to the eigenstates of the second subset, e.g., to excited electronic state and shifted vibrational state.
If the time of this transition is smaller than both inverse dephasing rate and inverse vibrational frequency, collapse of transition oscillations appear.
Due to the frequency difference between the Fock vibrational states and shifted Fock vibrational states are equal to each other, revival of transient oscillations appear.
These collapses and revivals proceed up to transient oscillations dissipate due to dephasing.
They manifest in the spectrum of polarization oscillations as splitting of the spectral line in the vicinity of electronic transition frequency.
Namely, spectral line splits into multiple lines with width equal to the inverse collapse time and distance between them equal to the vibron frequency.
The obtained results provides an additional tool for measurement of the vibron frequency and constant of interaction between electronic and nuclei vibration degrees of freedom.

\begin{acknowledgments}
The study was supported by a grant from Russian Science Foundation (project No. 20-72-10057). 
V.Yu.S. thanks foundation for the advancement of theoretical physics and mathematics "Basis".
\end{acknowledgments}

\bibliography{VibronDM}

\end{document}